\documentstyle[preprint,aps,psfig]{revtex}
\begin{document}

\title
{Clustering and signalling of cell receptors
}

\author{Yu Shi
\footnote{Email address: ys219@cam.ac.uk}
} 

\address{Cavendish Laboratory, University of Cambridge,
Cambridge CB3 0HE, United Kingdom}
\draft
\tighten

\maketitle

\begin{abstract}

As a response to ligand binding,
transmembrane cell receptors often enhance their clustering, or 
oligomerization, during the signalling process.
Here we present a statistical
mechanical model which combines the aspects of clustering and  signalling. 
In this model, receptors float on the surface, while 
for two neighboring  receptors, 
there is an interaction energy dependent on their conformational 
states. On the other hand, 
ligand binding of a receptor shifts the energy difference 
between the two conformational states. Due to   
thermal fluctuation,  the effects of
clustering and signalling  are statistical
average quantities. This model reduces to a floating Ising model with
a random field. We calculate the signalling in a grand
canonical ensemble mean field approach, using  Hubbard-Stratonovich 
transformation and replica method. Monte Carlo simulations are also performed. 
Essential biological features   are obtained in our model. 

\end{abstract}

\bigskip



\section*{1. Introduction}

On the surface of a cell, there are many transmembrane
receptor molecules. They are  ``sensory machines'' of the cell.
Binding   extracellular ligands changes the conformational states or, 
 in our point of view,  the probability distribution of 
the conformational states.    Consequently 
a cascade of responses occurs inside the cell.
This is called signalling, which is often quite sensitive.
An example is  bacterial  chemotaxis~\cite{berg}:
through the detection by its receptors, a bacterium  swims towards
higher concentration of a chemical it likes or a lower concentration
of a chemical it dislikes.  In this case, 
thousands of receptors cluster  together at a pole of the cell. 
Recently, based on  detailed analyses  on the experimental findings,
I proposed that cooperative  receptor-receptor
interaction  enhances the signalling, 
and constructed a statistical mechanical theory for this
process~\cite{shi1,shi2,shi3}. My idea has since received considerable
attention~\footnote{The reference was mostly made to a publication
which popularizes
some rudiments of my idea [T. Duke and D. Bray,
Proc. Nat. Acad. Sci. {\bf 96}, 10104 (1999)], and the attribution
of credit was inappropriately  made in T. Duke, N. Le Nov\`{e}re and
 D. Bray, J.~Mol.~Bio. {\bf 308}, 541 (2001).}.
In \cite{shi3},  good consistency was found 
in comparing this theory with  experimental
results~\cite{khan1,Khan,jasuja,spudich,Bornhorst,Sourjik}, and 
it was also suggested what kinds of experiment  are needed. 

In the case of many  other cell receptors, however,
usually they float  on the cell membrane. 
A ubiquitous phenomenon is that ligand 
binding enhances the  clustering of receptors 
during the signalling process.
Investigations on this issue  are widely reported for
various  receptors, for 
example, 
integins, which attach cells to the extracellular matrix, 
tyrosine kinase receptors, cytoskine receptors, growth factor 
receptors, tumor necrosis factor receptors, including death receptors, 
antigen receptors such as T-cell receptors, bone morphogenetic protein
receptor, G-protein-coupled receptors, 
etc~\cite{Lemmon1,Heldin,Lemmon2,Germain,Reich,Ashkenazi,Giancotti,Gilboa,Schulz}.  
In our previous works~\cite{shi1,shi2,shi3}, the position of 
each receptor is fixed, and there is an {\it a priori} cluster. 
In the present article, we shall consider statistical mechanics of 
floating receptors.  Based on our model, it is proposed that 
signalling sensitivity and clustering enhancement, in certain cases,  
are likely  two consequences of the same mechanism based on 
conformation-dependent  local receptor-receptor interaction.

In Sec.~2, we give a detailed description of the model. 
A mean field calculation of the signalling, as a statistical
average quantity quantity,
  is given in Sec.~3.  In Sec.~4, we discuss receptor 
clustering. Results of Monte Carlo simulations are 
given in Sec.~5.  Finally 
a summary and discussion is made in Sec.~6.

\section*{2. The Model}

Consider a collection of receptors on a lattice.
Fig.~1 is an illustrative configuration.  
Let the coordinate number be $\nu$, which is $6$ for a honeycomb lattice and 
is  $4$ for a square lattice.  Suppose that the receptors can move 
around on the lattice. Hence  at an instant, a receptor
has $\leq\nu$ nearest-neighboring receptors.
The conformational state  of the receptor  at lattice point 
$i$ is represented as  $V_i$, the value of which
is either $V^0$ or $V^1$, if we follow the two-state convention.
   $V_i$ can be understood
as the position of the receptor molecule.  Obviously,
$V_i$ is influenced by  whether this  receptor is bound to 
ligand.  We suggest  that $V_i$ is also  
influenced by  the conformational states  of its  
nearest-neighboring receptors, due to
 receptor-receptor interaction which is dependent on their conformational
 states.
We use 
$H_i$ to represent whether there is a ligand binding, that is,
$H_i=H\neq 0$ if the receptor is bound to ligand, otherwise $H_i=0$.
$H$ can be understood as the force or torque generated by 
the  ligand binding. $H V_i$ is the  energy  due to 
ligand binding.    Given the ambient  chemical concentration $[L]$,
the occupancy, i.e. the fraction of receptors with ligand bound, is
\begin{equation}
c=\frac{[L]}{[L]+K_d},
\end{equation}
 where  $K_d$ is the   dissociation constant.
Therefore each $H_i$  is   randomly distributed between $H$ and $0$, with 
probabilities $c$ and $1-c$,  respectively. 
Formally,  the probability distribution can be written as  
\begin{equation}
p(H_{i})\,=\,c\delta(H_{i}-H)+(1-c)\delta(H_{i}),\label{dis}
\end{equation}
where $\delta(x)=0$ if $x\neq 0$, while $\int \delta(x) dx=1$.
Since usually $K_d < [L]$, it is reasonable to regard the randomness of 
$H_i$ as quenched~\cite{shi2,shi3}.

We associate each site $i$ with a variable $n_i$, which is $1$ if there is 
a receptor on this site, and is $0$ otherwise.
In reality, of course $V_i$  and $H_i$ are 
only defined when there is a receptor
on site $i$.  However,  because the presence or not 
of a receptor at site  $i$ has been  determined  by $n_i$,
for the convenience of treatment,  equivalently 
we may associate   $V_i$ and $H_i$  to each site $i$, no matter whether
there is a  receptor on site  $i$.  
Since the  receptors are allowed to diffuse,
$n_i$ is time-dependent.  One may constrain that 
the total number of receptors, $N=\sum_in_i$,  is conserved. It
is also possible that the  receptors may get out of the relevant
region, hence $N$ is not conserved.
In any
case, we may use grand canonical ensemble approach characterized by
the chemical potential, which, in the first case, is given by the
constraint on the conservation of receptor number.

Therefore,  in absence of thermal fluctuation,  we have 
\begin{equation}
V_{i}\,=\,\psi(\sum_{j} t_{ij}n_j V_{j}+H_{i}-U_{i}),
 \mbox{with}\,
\psi(x) \,=\,\left\{  \begin{array}{ll}
V^{1} & \mbox{if $x>0$}\\
V^{0} & \mbox{if $x\leq 0$}
\end{array}\right. , \label{th}
\end{equation}
where the summation is over the nearest neighbors $j$ of site $i$,
 $U_{i}$ is
 a threshold value,  $T_{ij}$ is a coefficient for the coupling
between  nearest neighbors. 
$\sum_{j}t_{ij}n_j V_{j}$ is the influence, i.e.  force or torque,
 on the 
receptor at site $i$ from the nearest-neighboring sites.
$H_i$ is the influence, i.e. force  or torque,
generated by the ligand binding of 
the receptor at $i$ itself. Eq.~(\ref{th}) means that if the 
total influence on a receptor, 
as the sum of that due to  its ligand binding  
and that due to 
the interaction with the nearest-neighbors, is larger than a certain value
$U_i$, the receptors is in state $V^1$. Otherwise it is $V^0$.
Equivalently, with  $t_{ij}\,=\,t_{ji}$ and $t_{ii}\,=\,0$,
the dynamics  is determined in the following 
Hamiltonian, 
\begin{equation}
{\cal H}\,=\,-t\sum_{\langle ij\rangle}
 n_i n_j V_{i}V_{j}-\sum_{i}H_{i}n_iV_{i}
+U\sum_{i} n_iV_{i}, \label{hamiltonian} 
\end{equation}
where $\langle ij \rangle$  represents pairs of nearest neighbors, 
$t_{ij}$ and  $U_i$ have been  set to be  constants.
 Eq. (\ref{hamiltonian}) gives the total energy of the system, with
the kinetic energy  neglected.  The first term is the total 
interaction energy of nearest-neighboring pairs. The second term is the
energy due to ligand binding. The third term is the original energy, which,
together with the first term, 
determines the probability distribution of the conformational states
 of all the receptors
in the absence of ligand binding of any receptor.
 Without ligand binding and coupling with
others, the  energy of a  receptor is $UV^0$ for state $V^0$, and is
$UV^1$ for state $V^1$. With ligand binding but without
coupling, it is $(U-H)V^0$ for state $V^0$, and is
$(U-H)V^1$ for state $V^1$.   The interaction energy of the pair 
$\langle ij\rangle$, if sites $i$ and $j$ are both occupied by receptors, 
is $-tV_iV_j$, the value of which is dependent on both $V_i$ and $V_j$. 
Since ligand binding can cause  more $V_i$ to flip from $V^0$ to   $V^1$, it 
is required that $(H-U)(V^1-V^0) > 0$.  Precise form of ${\cal H}$
 is subject to
future experiments. However, (\ref{hamiltonian}) captures the essential 
elements while allows a theoretical treatment.

For convenience,
 we transform to a spin language, with the spin defined as
\begin{equation}
S_i=2\frac{V_i-V^0}{V^1-V^0}-1. \label{s}
\end{equation}
 Hence it is  $S_i=1$ if $V_i=V^1$, and 
is  $S_i=-1$ if $V_i=V^0$. In order to simplify the mathematics 
without loss of the essence, 
suppose $V^1=-V^0$.  Therefore  we obtain the following simplified
 Hamiltonian
\begin{equation}
{\cal H}\,=\,-J\sum_{<ij>} n_i n_j S_{i}S_{j}-\sum_{i}B_{i}n_iS_{i},
 \label{hamiltonian2} 
\end{equation}
 with $J=t(V^1-V^0)^2/4$, $B_i=(H_i-U)(V^1-V^0)/2$, which is thus 
$B_1=(H-U)(V^1-V^0)/2$ with 
probability $c$ and is $B_2=-U(V^1-V^0)/2$ with probability $1-c$,
 i.e.
\begin{equation}
p(B_{i})\,=\,c\delta(B_{i}-B_1)+(1-c)\delta(B_{i}-B_2). \label{dis2}
\end{equation}
We call this model, defined by 
(\ref{hamiltonian2}) and (\ref{dis2}),  floating
Ising model with a quenched random field  at each site. Note that it is
not a diluted Ising model, since $n_i$ is not a quenched disorder, but is
a dynamical variable on the same footing as $S_i$.  

It should be mentioned that a cell signalling process often 
involves different and largely separated time scales. On a much 
longer time scale the degrees of signalling and clustering evolve
after the stimulus, till 
they return to the pre-stimulus levels. So there is a piece of
short-time scale physics, and there is a piece of long-time scale physics. 
Here it suffices to discuss the  the short-time scale physics, i.e. the
quasi-equilibrium part of the theory. 
With an incorporation of the feedback, the long-time
scale behavior is similar to that described in Ref. \cite{shi2,shi3}.
With the adaptation process taken into account, the ligand concentration 
above should be replaced as the change of the ligand concentration. However, 
as studied in \cite{shi2,shi3},
perfect adaptation is equivalent to  no ligand binding.  

Guo and Levine studied the clustering  of receptors under the assumption  that 
the  interaction between neighbors depends on whether
the receptors are   liganded~\cite{Guo},
while the signalling problem  was not considered. As described in our 
model,  
We think that it is   more  likely   that the interaction depends on
the conformations  of the receptors, rather than 
whether the receptors are liganded.
In addition to the consistency with  
the situation of chemotaxis~\cite{shi1,shi2,shi3},
this is also supported by the strong  experimental evidence
that oligomerization is stabilized by
receptor-receptor interaction promoted by 
conformational changes of the 
receptors~\cite{Lemmon1,Heldin,Lemmon2,Germain,Gilboa,Schulz}.

\section*{3. Mean-field calculation of signalling}

The activity of the system is $\langle \sum_i n_i V_i\rangle$, where
the summation is over all the lattice sites,
$\langle \dots \rangle$ represents ensemble average, i.e. average over
the probability distribution of all possible configurations of the
system. It is straightforward to write it as $(V^1-V^0)W/2$, with 
$W=\langle  \sum_i n_iS_i\rangle=Im$, where  $m= \langle
n_iS_i\rangle$ is the average spin per site, I is the number of
lattice sites.  
The signalling can be characterized by 
the {\em change} of $W$  when the
occupancy is changed from $c=0$ to $c$,
i.e. $W(c)-W(c=0)=I[m(c)-m(c=0)]$.

Therefore we set out to calculate $m$ as a function of $c$,
using grand canonical ensemble approach.
The grand partition function is
\begin{equation}
{\cal Z}
=\prod_{i}(\sum_{n_i=0,1}\sum_{S_i=-1,1})(\prod_{i}e^{\beta\mu \sum_i n_i})
e^{-\beta{\cal H}},
\end{equation}
where $\mu$ is the chemical potential, ${\cal H}$ is as given in 
(\ref{hamiltonian2}).
Using Hubbard-Stratonovich transformation, we obtain
$$
e^{-\beta{\cal H}}=
A\int (\prod_i d\phi_i)
\exp[-\frac{1}{2}\sum_{i,j} \phi_i K^{-1}_{ij}\phi_j+
\sum_{i}(\phi_i+\beta B_i)n_iS_i]$$
where $A=(2\pi)^{-I/2}(det K)^{-1/2}$. $K=\beta(J_{ij})$ is a matrix,
$K^{-1}$ is its inverse.  Consequently
\begin{equation}
{\cal Z}(\{B_i\})=A\int (\prod_i d\phi_i)
\exp(-\frac{1}{2}\sum_{i,j} \phi_i K^{-1}_{ij}\phi_j)
\prod_{i}[2+e^{\beta\mu+\phi_i+\beta B_i}+e^{\beta\mu-\phi_i-\beta B_i}].
\end{equation}
Based on 
%
$\ln {\cal Z} = \lim_{n \rightarrow 0}({\cal Z}^n-1)/n$,
the replica method yields
\begin{equation}
\ln {\cal Z}(\{B_i\})
= \lim_{n\rightarrow 0}\frac{1}{n}\left\{
A^n\int(\prod_{\alpha,i}d\phi_i^{\alpha})
\exp (-\frac{1}{2}\sum_{\alpha,i,j} \phi_i^{\alpha} K^{-1}_{ij}\phi_j^{\alpha})
\prod_{i,\alpha}
[2+e^{\beta\mu+\phi_i^{\alpha}+\beta B_i}+
e^{\beta\mu-\phi_i^{\alpha}-\beta B_i}]-1\right\}, \label{zb}
\end{equation}
where $\alpha=1,\cdots,n$ denote the replicas.

Averaging $\ln {\cal Z}$ over the $B_i$ distribution   (\ref{dis2}),
one obtains
\begin{eqnarray}
&&\overline{\ln {\cal Z}} = \int \ln {\cal Z}(\{B_i\}) \prod_i p(B_i) \prod_i
dB_i \nonumber \\
&&=\lim_{n\rightarrow 0}\frac{1}{n}\big\{ A^n
\int(\prod_{\alpha,i}d\phi_i^{\alpha})
\exp (-\frac{1}{2}\sum_{\alpha,i,j} \phi_i^{\alpha} K^{-1}_{ij}\phi_j^{\alpha})
\nonumber\\
&& \times\prod_{i}\left[ 
c_1\prod_{\alpha}(2+e^{\beta\mu+(\phi_i^{\alpha}+\beta B_1)}
+e^{\beta\mu-(\phi_i^{\alpha}+\beta B_1)})
+c_2\prod_{\alpha}(2+e^{\beta\mu+(\phi_i^{\alpha}+\beta B_2)}
+e^{\beta\mu-(\phi_i^{\alpha}+\beta B_2)})]\right]-1\big\}, \label{zbs}
\end{eqnarray}
where, for notation convenience, we have  set  $c_1=c$, and $c_2=1-c$.

The mean field value of $\phi^{\alpha}_i$ can
be obtained by deciding 
the saddle point of the integrand in Eq.~(\ref{zbs}), for which one can obtain
\begin{equation}
\sum_j K_{ij}^{-1}\Phi_n
 =  \frac{c_1 f_{1}^{n-1}g_1+c_2f_2^{n-1}g_2}
{c_1f_1^n+c_2f_2^n},
\end{equation}
where $\Phi_n$ is the saddle point value of  $\phi^{\alpha}_i$, corresponding
to $n$ replicas,   
$f_{k}=2+e^{\beta\mu+(\Phi_n+\beta B_k)}+
e^{\beta\mu-(\Phi_n+\beta B_k)}$, $g_{k}=e^{\beta\mu+(\Phi_n +\beta B_k)}-
e^{\beta\mu-(\Phi_n+\beta B_k)}$, ($k=1,2$). Taking $n \rightarrow 0$ limit,
we have 
$\sum_j K_{ij}^{-1} \Phi = c_1g_1/\overline{f}_1+
c_2 g_2/\overline{f}_2$,
where $\Phi= \lim_{n\rightarrow 0}\Phi_n$, 
$\overline{f}_{k}=2+e^{\beta\mu+(\Phi+\beta B_k)}+
e^{\beta\mu-(\Phi+\beta B_k)}$, 
$\overline{g}_{k}=e^{\beta\mu+(\Phi+\beta B_k)}-
e^{\beta\mu-(\Phi+\beta B_k)}$, ($k=1,2$). Therefore
\begin{equation}
\Phi = \beta\nu J \left( c_1\frac{g_1}{\overline{f}_1}
+c_2\frac{g_2}{\overline{f}_2}\right). \label{pp}
\end{equation}
is the mean-field value of the Hubbard-Stratonovich field. 
The saddle point approximation implies that 
\begin{equation}
\overline{\ln {\cal Z}} = \lim_{n\rightarrow 0}\frac{A^n}{n}
\left\{\exp(-\frac{n}{2}\sum_{i,j} \Phi K^{-1}_{ij}\Phi)
(c_1\overline{f}_1^n+c_2\overline{f}_2^n)^{I}-1\right\}.
\end{equation}

 $m$, as a measure of the activity, is calculated
in the following way. For given $\{B_i\}$, the ensemble average of $S_i$ is
$\langle S_i \rangle_{\{B_i\}}=\partial lnZ(\{B_i\})/
\partial(\beta B_i)
= \partial lnZ(\{B_i\})/\partial(\phi_i)
= \sum_j  K^{-1}_{ij} \langle \phi_i\rangle_{\{B_i\}}$,
whose average over $p(\{\{B_i\})$ gives the  $m$. Thus
$m=\overline{\langle S_i\rangle_{\{B_i\}}}= \sum_j  K^{-1}_{ij} \Phi$,
where $\Phi$ is the mean-field value of the 
 Hubbard-Stratonovich field, as given in (\ref{pp}).
Therefore $\Phi= \beta\nu J m$. 
 Eq.~(\ref{pp}) can thus be written as
\begin{equation}
m=\sum_{k=1,2} c_k \frac{e^{\beta\mu+\beta\nu Jm+\beta B_k}
-e^{\beta\mu-\beta\nu Jm-\beta B_k}}{2+e^{\beta\mu+\beta\nu Jm+\beta B_k}
+e^{\beta\mu-\beta\nu Jm-\beta B_k}}. \label{m}
\end{equation}

The average receptor number is
$\langle N\rangle
=\partial \overline{\ln {\cal Z}}/\partial(\beta\mu)$, and can 
be evaluated as
\begin{equation}
\langle N \rangle =I\sum_{k=1,2} c_k \frac{e^{\beta\mu+\beta\nu Jm+\beta B_k}
+e^{\beta\mu-\beta\nu Jm-\beta B_k}}{2+e^{\beta\mu+\beta\nu Jm+\beta B_k}
+e^{\beta\mu-\beta\nu Jm-\beta B_k}}. \label{n}
\end{equation}

(\ref{m}) and (\ref{n})  can also be obtained in a simple mean field approach
regarding $H-\mu N$ as an effective Hamiltonian, and considering four
possibilities of the state $(n_i, S_i)$ at site $i$, subject to an
effective field $\nu Jm+B_i$. 

Hence we have obtained  an analytical expression for 
the signalling, as a statistical average quantity. If the receptor
number is conserved
in the system, (\ref{n}) gives $\mu$ as a function of the receptor number. 

Sensitivity  of the signalling to ligand
concentration is given by 
$\partial m/\partial c$, which can be arbitrarily large. 

For annealed randomness, which may apply to the rare  case  in which $[L]$
is comparable to $K_d$, one may obtain
\begin{equation}
m=\frac{\sum_{k=1,2}c_k(e^{\beta\mu+\beta\nu Jm+\beta B_k}
-e^{\beta\mu-\beta\nu Jm-\beta B_k})}
{\sum_{k=1,2}c_k(2+e^{\beta\mu+\beta\nu Jm+\beta B_k}
+e^{\beta\mu-\beta\nu Jm-\beta B_k})}, \label{m2}
\end{equation}
and 
\begin{equation}
\langle N\rangle =I\frac{\sum_{k=1,2}c_k(e^{\beta\mu+\beta\nu Jm+\beta B_k}
+e^{\beta\mu-\beta\nu Jm-\beta B_k})}
{\sum_{k=1,2}c_k(2+e^{\beta\mu+\beta\nu Jm+\beta B_k}
+e^{\beta\mu-\beta\nu Jm-\beta B_k})}, \label{n2}
\end{equation}

In the  high temperature limit, the difference between (\ref{m}) and
(\ref{m2}), and that between (\ref{n}) and  (\ref{n2}) tend to diminish. 

\section*{4. clustering}

The conformation-dependent interaction  not only enhances signalling 
sensitivity, it is  also responsible for clustering.
The first term of the Hamiltonian  (\ref{hamiltonian2}) implies 
that, in order to minimize the Hamiltonian, 
the receptors tend to aggregate  together to maximize the
number of nonzero 
$n_in_j$ with $S_i=S_j$ for neighboring $\langle ij\rangle$.
Therefore receptors with the same conformational state
 tend to cluster. However, because of  thermal fluctuation, 
they cannot all  cluster 
together, since there is a probability distribution for various 
configurations.  As an illustration of
the situation, a snapshot  obtained in a  Monte Carlo
simulation  is  shown in Fig. 1.

The second term in  (\ref{hamiltonian2}) determines how ligand binding
affects the clustering situation. 
It can be seen that    $U$ 
 determines the bias of the distribution of the 
receptor state $\{ S_i \}$ between $1$ and $-1$
when there is no ligand binding ($H_i=0$).
 That is, if $U=0$, for not-too-strong interaction  compared with the 
temperature ($\beta J$)\footnote{There is no such  a limitation  if
the adaptation on a long time scale is considered, 
see Ref.~\cite{shi2,shi3}.}, 
there is an equal number of 
receptors with spin $1$ and those with $-1$.  If $U>0$, there are more 
receptors with spin $-1$. $U$ also
determines whether ligand binding enhances or suppresses the clustering. 
For example, we may consider two typical  cases:

Case (i), $U=H$. According to (\ref{th}), this
 means that merely a ligand  binding of a receptor
is enough to change its conformational  state.
In this case,  $B_1=0$, $B_2=-H(V^1-V^0)/2$.

Case (ii), $U=0$. According to (\ref{hamiltonian}), 
this means that in the absence of ligand binding ($H_i=0$),
there is no bias in the state distribution. In this case, 
 $B_1=H(V^1-V^0)/2$,  $B_2=0$.

These two cases can map to each other with the transformation 
$H\rightarrow -H$ and $c\rightarrow 1-c$.

In case (i), since $B_2<0$, in the absence of ligand, all $S_i$ tend to be
close to $-1$. Ligand binding causes more  $S_i$ to be $1$. Because 
receptors with the same value of $S_i$ tend  to cluster, those  
with different $S_i$ tend  to be disconnected, the conclusion is that
ligand binding suppresses clustering. 

In case (ii), if there is no  
ligand binding, there is no biasing field, so there
are equal probabilities  for $S_i$ to be $1$ and $-1$, consequently 
the clustering is minimized.  When there is ligand binding, the clustering
is enhanced. 

For other values of $U$, i.e. $U\neq 0$ and $U\neq H$,
we have the following general picture.
Without ligand binding, there is a uniform field $B_2$ at every site. 
In the presence of ligand binding, there is a random field as given by
(\ref{dis2}). 
With a rough mean field estimation, one may see that 
whether the clustering is enhanced depends on whether $(1-c)B_2+cB_1$
has larger absolute value than  $|B_2|$. Thus    
only if $H>2U/c$, is the clustering enhanced by ligand binding. 
Case (ii) belongs to  this region.  If $U=H/2$, we have $B_1=-B_2$.
 
The clustering of receptors can be studied quantitatively  by 
defining a clustering correlation function as
\begin{equation}
C(r)=\frac{\langle n_i n_{i+\bbox{r}}\rangle-\langle n\rangle^2}
{\langle n\rangle ^2},
\end{equation}
where the average is over different sites  $i$ and 
different directions of $\bbox{r}$ first,
and then over the thermodynamic 
 ensemble, i.e.  different possible configurations
at the same temperature. 
$\langle n\rangle = \langle N \rangle/I$
 is the density of receptors on the lattice.
If there is no clustering correlation, $\langle n_i n_{i+\bbox{r}}\rangle$$=
$$\langle n\rangle^2$, consequently  $C(r)=0$. This correlation function 
measures the deviation from the non-correlating case and allows comparison
of situations for different receptor densities.  This definition is
similar to the two-point correlation function in the study of galaxy
clustering.  One may also define higher-order correlation functions.
We  leave for future investigations 
the analytical calculation of  the clustering correlation function,
 and turn to numerical simulations here. 
 
\section*{5. Results of Monte Carlo simulations}

  To investigate  the extent of clustering and to calculate the activity,
  we have done  Monte Carlo
simulations using the  Metropolis algorithm on a square lattice. 
In the simulations, we conserve the receptor numbers.  
We have specifically studied Case (ii),  from which
 one may obtain the results for
case (i)  simply by changing $c$ to $1-c$.
 The results show  strong 
correlations for small values of $r$ and weak correlations for large
  values
of $r$. The decay looks like exponential, as expected for a
  non-critical system.

We studied correlation functions for different values of the 
 ligand binding fraction $c$, with a same receptor density
$\langle n \rangle$. See Figs.~2(a) and 2(b) for results under different values
of the coupling-noise-ratio $\beta J$. 
From the correlation function for small values of $r$,
it is clear that the larger the ligand binding fraction 
$c$,
the larger the correlation. This  confirms the
above analyses. 

Comparing Figs. 2(a) and 2(b), it can be seen  that 
 with larger coupling-noise ratio $\beta J$,
 as in Fig~2(a), the correlation function $C(r)$
is larger at small values of $r$, while smaller at large values of
  $r$,   indicating that  clustering is  stronger. On the other hand,
  for larger $\beta J$, the decay of $C(r)$ with $r$ is faster,
  because it is further from criticality.

We also studied the correlation functions for different values of 
 receptor densities $\langle n\rangle$ with 
a same ligand binding fraction $c$ (see Fig. 2(c)). 
It is shown that the smaller the receptor density, the larger the 
correlation function.
This can be understood, since the larger the density,
the less freedom two receptors can approach each other. 

In the simulations, we   also  obtained  the activity.
The activity $W$ can also be written as $\sum_i' S_i$, where
$\sum_i'$ represents  summation over the receptors instead of the
lattice sites. Hence  it equals $NM$, where $M$ is the
average activity    per receptor. 

Fig. 3 gives the  relation between
the average
activity per receptor $M$  and the ligand binding occupancy $c$, for different
values of coupling-noise ratio $\beta J$. 
 From the plots, we see that the activity increases with the
ligand binding fraction
 $c$, with  $\beta J$, 
and with the receptor density $\langle n\rangle$, as consistent with
the mean field solution.
  
\section*{6. Summary and Discussion}

In this paper, we propose a statistical mechanical theory which accounts for
 clustering and signalling of receptors in a same framework. 
 Clustering and signalling of  a network of receptors
are treated as statistical average quantities.
In our model, which can be reduced
  to a so-called  floating Ising model in a random field, 
the interaction energy between  neighboring receptors depends on their
conformational states, therefore through this interaction, the 
conformational 
state of one receptor influences those  of its  neighbors. 
On the other hand, since the receptors are allowed 
 to move on the membrane, the
receptors with a same conformational state tend to cluster together,
in order to decrease the total energy of the system. Therefore 
   clustering and signalling are unified as two consequences of the 
same coupling between receptors. 

According to our theory, clustering exists even in absence of the
ligand binding.
This is  consistent  with a  recent experimental 
finding~\cite{Gilboa}.
We have studied  clustering, based on an appropriate definition
of  clustering correlation function. Monte Carlo simulations were made. 
To obtain  the situation that ligand binding enhances the clustering,
a  parameter, namely 
 the threshold value $U$ for  the change of 
the conformational state, which determines 
 the state distribution in the absence
 of ligand binding, must be within a certain range. 
If the values of the variable
  characterizing the two conformational states  are 
symmetric, i.e. with a same magnitude and with opposite signs, then 
a simple possibility is that this threshold value is zero.
 We note that
it was found experimentally that, in the absence of ligands, 
the receptors are hindered to cluster by certain inhibitors, which
are squeezed out when ligands bind~\cite{Jiang}.
In such a case, it may be constrained  that
clustering is always enhanced by ligand binding.

By using a mean field theory based on Hubbard-Stratonovich transformation 
and replica method, as well as by using Monte Carlo simulations, we also 
studied the activity as a function of ligand binding fraction $c$.

This model is  also be interesting from the point of the view of statistical 
mechanics.  On the other hand, it
 is straightforward to make appropriate extensions of this model, for
 example,
to put in more details of the realistic systems.
The adaptation can be studied by straightforwardly 
generalizing our  previous approach
based on a counteracting field as a feedback from the
signal to the field~\cite{shi2}.
In this case, both signalling and
 clustering are adapted through a feedback, on a long time scale. 

A simple and direct experimental test of our theory  is to examine whether 
conformational state of a receptor can be changed by ligand binding of its 
nearest-neighboring receptors. It is interesting 
to study the forces generated by
ligand bindings of a receptor and its neighbors.   
Parameters in our model need to be measured.  The techniques 
used in studying single molecules may be useful for this subject. 

\section*{acknowledgement}

I am grateful to Prof. Howard Berg at Harvard University for enlightening,
useful and pleasant  discussions during 
 my work on  Ref.~\cite{shi3} and the current paper. 

\newpage

Figure Captions:

Fig 1. A configurational snapshot of 300 receptors
on a $50\times 50$ lattice.  It  is in
$10000$  Monte Carlo steps after an initial random configuration.
An empty circle represents a receptor with $S_i=-1$ and $B_i=0$,
a filled  circle represents a receptor with $S_i=-1$ and $B_i=B$,
an empty square  represents a receptor with $S_i=1$ and $B_i=0$,
a filled  square represents a receptor with $S_i=1$ and $B_i=B$.
The probability for $B_i$ to be $B$ is $0.8$. $\beta B=2$, $\beta J=0.8$. 
Obviously, increasing $\beta J$ enhances clustering.

Fig 2. Monte Carlo calculation of
the clustering correlation function
$C(r)$ for receptors on  a $50\times 50$ lattice.
The result is obtained by averaging  over different 
sites with the same distance, and over  $1000$ Monte Carlo 
steps following $10000$ initial steps to approach equilibrium.  
$\beta B=2$.  (a) $\beta J=1.6$,
there are 300 receptors,
results for different values of ligand binding fraction
are displayed as different symbols. 
(b) $\beta J=0.8$,
there are 300 receptors, 
results for different values of ligand binding fraction  
are displayed as different symbols. 
(c)  $\beta J=0.8$,
the ligand binding fraction  $c=0.8$,
results for different total numbers of receptors are displayed as 
different symbols.

Fig3. Monte Carlo calculation of
the activity  of the receptors on a $50\times 50$ lattice.
For comparison between different densities of receptors, 
we give  the relation between
 the average ``spin'' per receptor $M$ and the  
ligand binding occupancy $c$.
The result is obtained by averaging  over all  receptors 
and over  $1000$ Monte Carlo steps following
$10000$ initial steps to approach equilibrium.  $\beta B=2$.
 (i) $N=300$, $\beta J=0$ ($\bullet$), 
i.e. there is no receptor-receptor interaction.
(ii) $N=300$, $\beta J=0.4$ (+). (iii) $N=300$, 
$\beta J=0.8$ ($\times$). (iv)  $N=300$, $\beta J=1.6$ 
($\odot$). In this case, the deviation of $M(c=0)$ from $0$ is  
 spontaneous ``magnetization'' due to strong coupling,  but it is always
brought to $0$ by  the
negative feedback on a long  time scale~\cite{shi2,shi3}. 
(v) $N=900$, $\beta J=0.8$ ($\diamondsuit$).

\psfig{figure=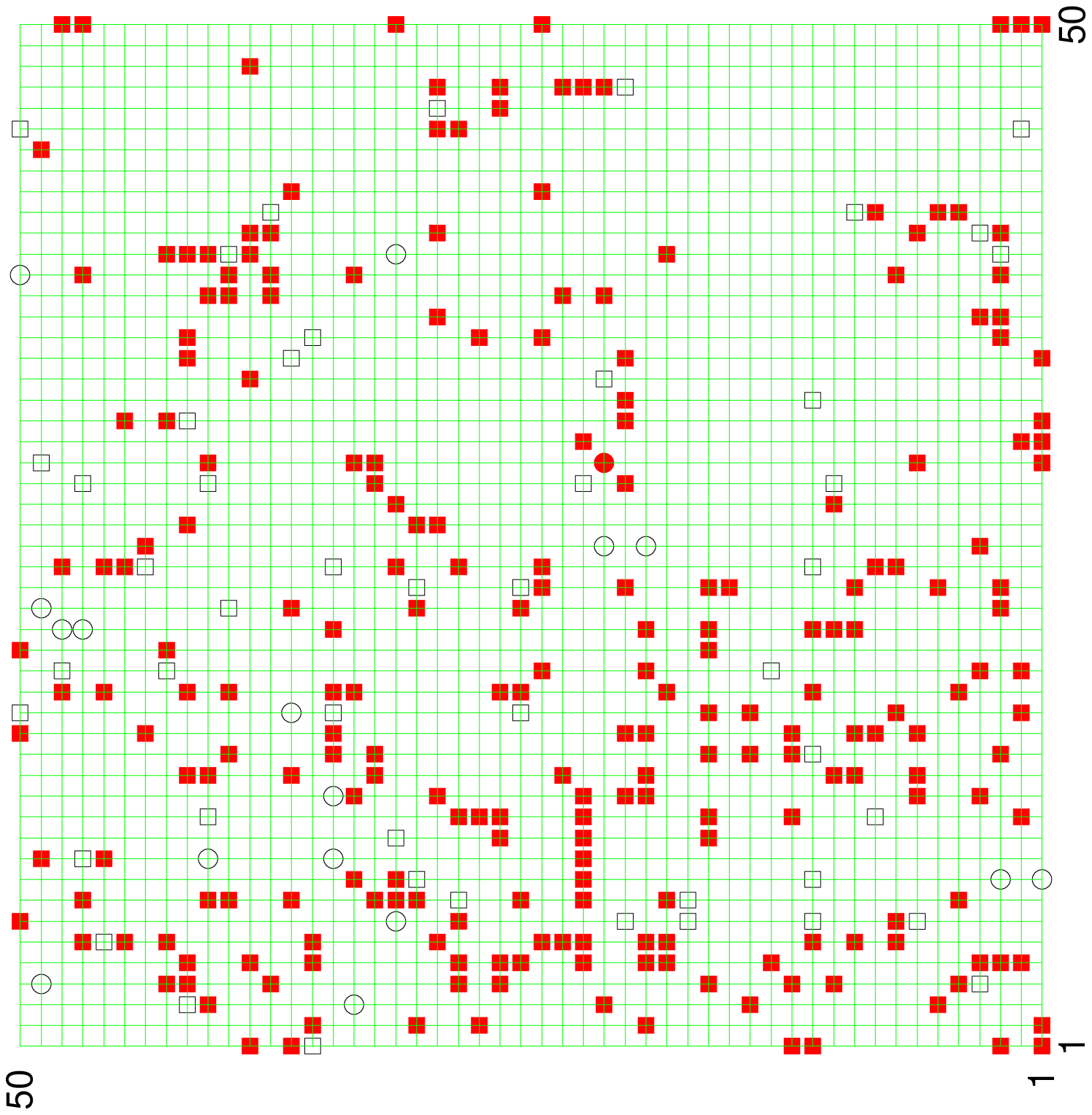}
\psfig{figure=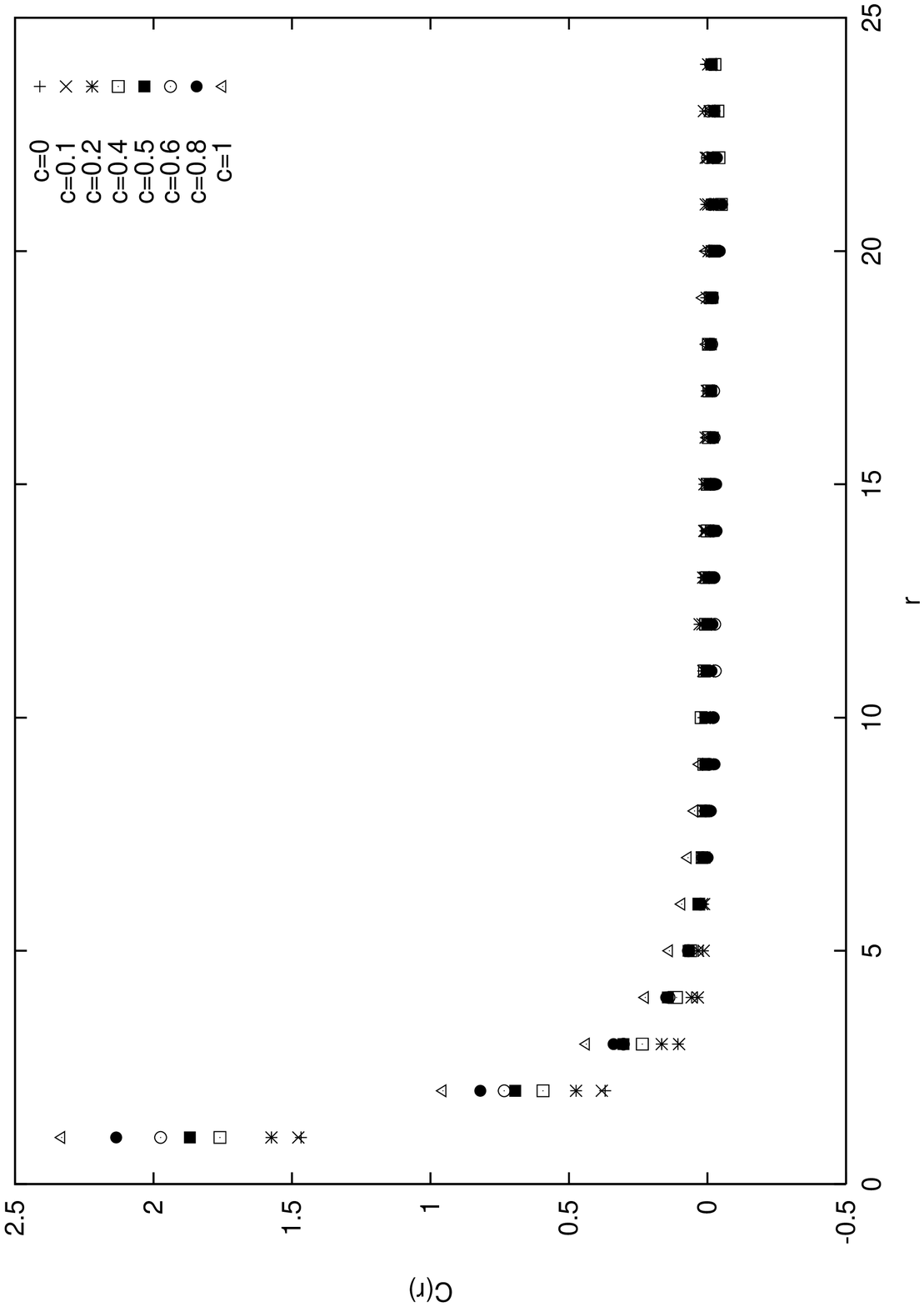}
\psfig{figure=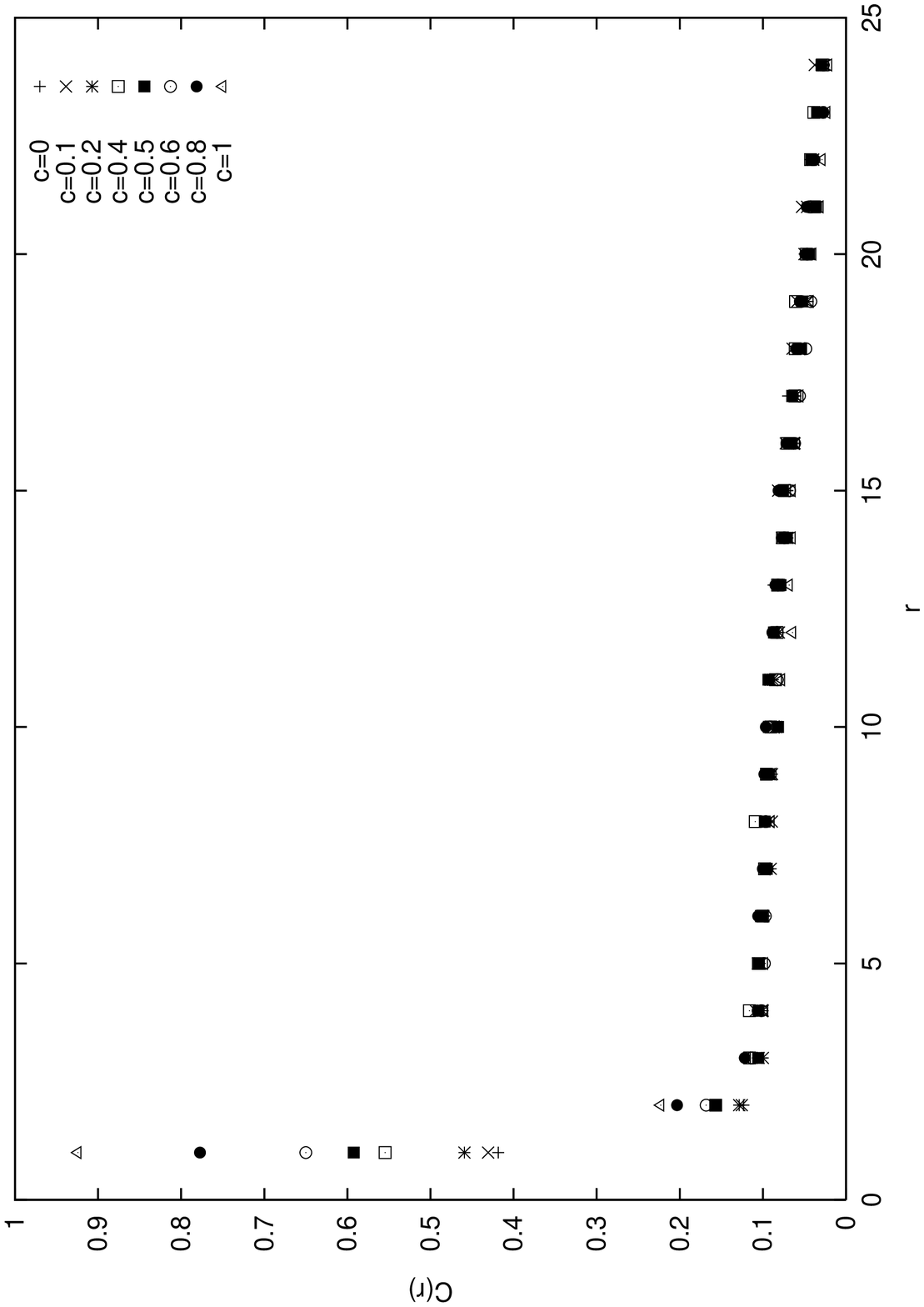}
\psfig{figure=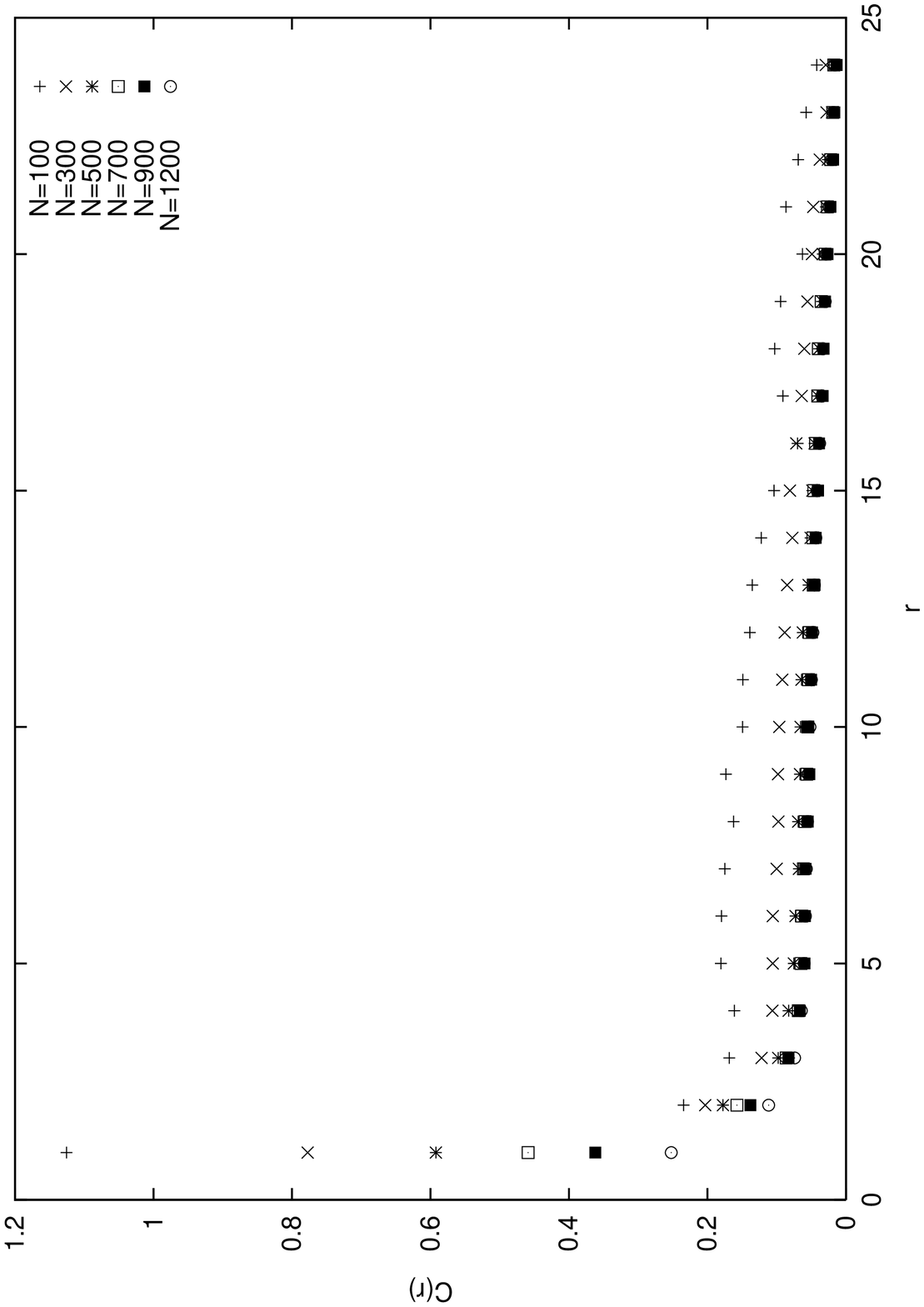}
\psfig{figure=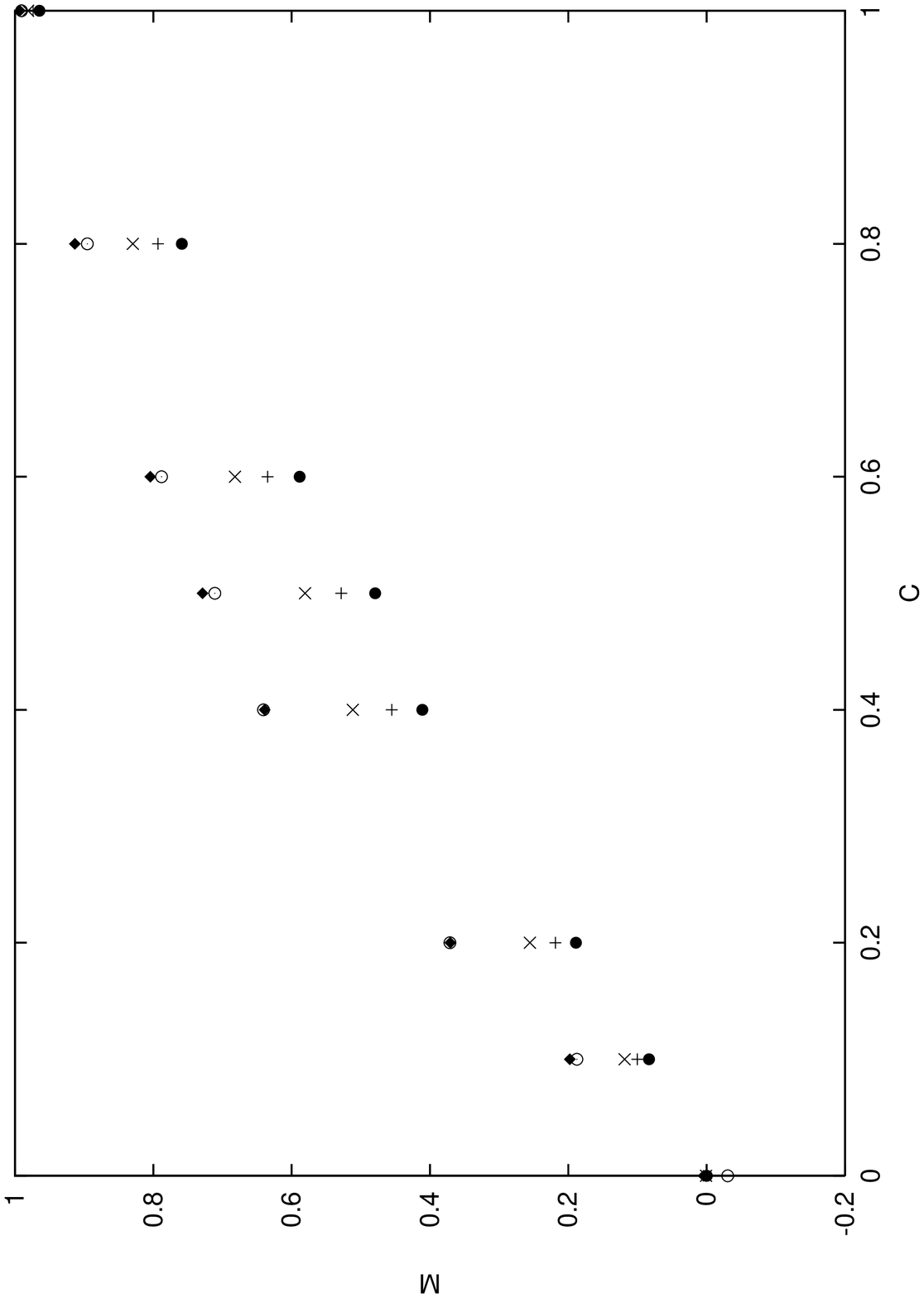}

\end{document}